\documentclass[amsmath,amssymb,showpacs]{revtex4-1}

\usepackage{graphicx}
\usepackage{natbib}
\usepackage{bm}
\usepackage{url}
\usepackage{textcase}
\usepackage{epsfig}

\begin{document}

\title{Chaplygin Gas of Tachyon Nature Imposed by Noether Symmetry and Constrained via 
$H(z)$ Data}

\author{L. G. Collodel}
 \email{lgcollodel@gmail.com}
\author{G. M. Kremer}
 \email{kremer@fisica.ufpr.br}
\affiliation{
 Departamento de F\'{i}sica, Universidade Federal do Paran\'{a}, 81531-980 Curitiba, Brazil}
\date{} 

\begin{abstract}
An action of general form is proposed for a Universe containing matter, 
radiation and dark energy. The latter is interpreted as a tachyon field 
non-minimally coupled to the scalar curvature. The Palatini approach is used 
when varying the action so the connection is given by a more generic form. Both 
the self-interaction potential and the non-minimally coupling function are 
obtained by constraining the system to present invariability under global point 
transformation of the fields (Noether Symmetry). The only possible solution is 
shown to be that of minimal coupling and constant potential (Chaplygin gas). The 
behavior of the dynamical properties of the system is compared to recent 
observational data, which infers that the tachyon field must indeed be 
dynamical.
\end{abstract}

\maketitle

\section{Introduction}
Tachyons have been vastly studied in M/String theories. Since the realization of 
its condensation proprieties, researchers have gained interest in its 
applications in cosmology. 
At first, there was the problem of describing the string theory tachyon by an 
effective field theory that would lead to the correct lagrangian in classical 
gravity. The first classical description of the tachyon field (\cite{sen,sen1}) addressed the lagrangian problem, making way for building the first model within 
tachyon cosmology (\cite{gibbons}).

Being a special kind of a scalar field, it present negative pressure, making the 
tachyon a natural candidate to explain dark energy (\cite{C,C1,C2,C3}). The 
inflationary period 
could also be explained if one considers the inflaton to behave as a tachyon 
field. Many different attempts were made under this assumption, testing a wide 
variety of self-interacting 
potentials such as power-laws, exponentials and hyperbolic functions of the 
field (\cite{D,D1,D2,D3,D4,D5,D6,D7}). The possible case scenario where the 
tachyon plays both roles, 
inflaton and dark energy, 
has also been studied in the works (\cite{E,E1}), where the first establishes 
constraints on the potential so the radiation's era could commence.

The studies above introduced a tachyon field which is minimally coupled through 
the metric, hence providing just another source for the gravitational field. 
Nevertheless, such fields 
might also be considered to be non-minimally coupled to the scalar curvature, 
becoming part of the spacetime geometry by generating a new degree of freedom 
for gravity. In this 
scenario, the gravitational constant $G$ becomes a variable function of 
spacetime.

Tachyon fields in the non-minimal coupling context were analyzed for both the 
inflationary period (\cite{G}) and the current era (\cite{H}). 
In those cases, the coupling functions and the self-interacting potentials were 
given in an \emph{ad-hoc} manner, as exponentials and power-law forms.

Every time we choose a different coupling or potential function, we create a new 
cosmological model, or even a new theory of gravity in the non-minimal case. 
This is a very difficult 
task since the lack of experiments and observations obligates one to find 
heuristics arguments to support the choice made. The advantages of searching for 
symmetries in systems where 
the lagrangian is known is widely entertained, it not only helps us find exact 
solutions but might also give us physically meaningful constants of movement. 
What is less appreciated is
the fact that one can constrain a system (one that lacks a closed form of the 
functional) to present symmetry. In what concerns non-minimally coupled tachyon 
fields, Noether symmetries
were used to establish the coupling and self-interaction functions in the 
papers (\cite{K,LK}). The latter makes use of the Palatini approach, in a way 
to generalize the theory, since
the non-minimal coupling can provide a metric-independent connection.

The Chaplygin gas was first introduced by Chaplygin in 1902 (\cite{chaplygin}) 
in the realms of aerodynamics. This gas features an exotic equation of state 
($p_c\propto-1/\rho_c$), 
which was originally used to describe the lifting force on a wing of an 
airplane. Because its pressure is negative, the Chaplygin gas became a good candidate to explain 
dark energy (\cite{kamenshchik,fabris,bilic,chaplygindm,kamenshchik4,kremerchaplygin}). The 
attempts to correlate fields and fluids soon showed that the constant potential 
tachyon field behaves as 
a Chaplygin gas (\cite{frolov,kamenshchik3,chimento,delcampo}). Its equation of state 
allows generalizations, giving rise to the so called \emph{Generalized Chaplygin 
Gas}, or just GCG. This gas exerts a negative pressure proportional in moduli to 
the inverse of some 
power of its energy density and was investigated in works such 
as (\cite{mbiesiada,fywan,zhzhu,lxujlu,ywang}), including its relationship to 
a - now, not constant potential - tachyon 
field (\cite{ggupta}). Originally, the equation of state of a Chaplyigin gas 
was so simple that even with the exhausted studies about the GCG there was still 
plenty of room for 
further generalizations. Endowing the EoS with a linear barotropic term, which 
alone would describe an ordinary fluid, enriched the GCG which under this 
assumption is called the 
\emph{Modified Chaplygin Gas}, MCG. Its motivations lie 
precisily on the possible field nature of the gas (\cite{hbbenaoum}), and its 
parameters have been constrained via observational analysis (\cite{bcpaul}). 
Further generalizations account for higher order energy density terms in the EoS 
of a Chaplygin Gas, the \emph{Extended Chaplyigin Gas}, 
ECG (\cite{bpourhassan,jlu}).

In this work, we start from a very general lagrangian for a tachyon field 
non-minimally coupled to the scalar curvature. Matter and radiation fields are 
also included in the system as
perfect fluids from the beginning. The connection is initially taken to be 
metric independent and the action is also varied with respect to it, a process 
known as the Palatini 
approach. Since we consider a flat, homogeneous and isotropic Universe, the flat 
Friedmann-Lema\^{i}tre-Robertson-Walker (FLRW) metric is used to rewrite our 
functional in the 
form of a point-like lagrangian. This presents an extra term than usual, which 
comes from the independent connection. The system is constrained to that one 
which presents invariance 
under continuous point transformations, or a Noether symmetry. The coupling and 
self-interaction potential functions of the tachyon field are then determined. 
Every new field added 
to the lagrangian clearly influences these point transformations. For that 
matter, it is important to start off from a complete system (including the 
radiation fields) if one takes 
symmetry as a first principle. We show that for this system to be Noether 
symmetrical, the non-minimal coupling must vanish and the self-interaction 
potential must be constant, hence the tachyonic Chaplygin gas.  The system is initially composed of five free parameters, namely the Hubble constant, the three density parameters for recent times and the normalized constant potential. The radiation parameter is then established in a \emph{ad-hoc} way so we are left with four different free parameters. These are determined via the $\chi^2$ analysis for the recent $H(z)$ data from SNe and gamma-ray bursts. We show that although dark energy tends asymptotically to a cosmological constant, any small discrepancies make the tachyon field dynamically active, so the Chaplygin gas presents property of transition from pressureless matter to dark energy.

In order to clarify the typos and notations here used, we remark: the metric 
signature is ($+,-,-,-$); the Levi-Civita connection is written with a tilde, 
$\tilde{\Gamma}^{\lambda}_{\mu\nu}=\left\{^{\lambda}_{\mu\nu}\right\}$ while the 
independent connection is given without it $\Gamma^{\lambda}_{\mu\nu}$. 
Natural constants were rescaled to the unity ($8\pi G=c=1$). Throughout the 
whole paper, derivatives in equations are presented as follows: dots represent 
time derivatives, while $\partial_{q^i}\equiv\frac{\partial}{\partial q^i}$ and 
$\partial_{\dot{q}^i}\equiv\frac{\partial}{\partial\dot{q}^i}$ stand for partial 
derivatives with respect to the generalized coordinate $q^i$ and velocity 
$\dot{q}^i$, respectively. 
\section{Action and Point Lagragian}

A generalization of the general theory of relativity is proposed through a 
non-minimal coupling of a function of the tachyon field. The general action for 
both geometry and source is written
\begin{equation}
\label{action}
S=\int 
d^4x\sqrt{-g}f(\phi)R-V(\phi)\sqrt{1-\partial_{\mu}\phi\partial^{\mu}\phi}
-\mathcal{L}_s,
\end{equation}
where $\phi$ is the tachyon field, $f(\phi)$ is the non-minimal coupling 
function, $V(\phi)$ is the self-interaction potential and $\mathcal{L}_s$ is the 
lagrangian density of other sources (matter and radiation).

In order to attain a more general theory we allow the connection to be metric 
independent. The variation of the action with respect to the connection 
$\Gamma^{\rho}_{\mu\nu}$ results in the well-known form 
\begin{equation}
\label{connection}
\Gamma^{\rho}_{\mu\nu}=\tilde{\Gamma}^{\rho}_{\mu\nu}+\frac{1}{2f}\left(\delta^{
\rho}_{\nu}\partial_{\mu}f+\delta^{\rho}_{\mu}\partial_{\nu}f-g_{\mu\nu}
\partial^{\rho}f\right).
\end{equation}
where $\tilde{\Gamma}^{\rho}_{\mu\nu}$ is the Levi-Civita connection.

Usually, the self-interaction potential and the coupling function are set in a 
\emph{ad-hoc} manner. Instead of approaching the problem this way, we would like 
to constrain the system to that which has a Noether symmetry. This is done by 
operating a variational vector field on the point-like lagrangian, and for this, 
we need to rewrite it on a specific metric. For a flat, homogeneous and 
isotropic Universe, spacetime is described by the flat FLRW metric. The 
point-like functional in (\ref{action}) then becomes
\begin{eqnarray}
\label{pll}
L=6f(\ddot{a}a^2+\dot{a}^2a)-\frac{3a^3}{2f}(\partial_{\phi}f\dot{\phi})^2+3a^3\partial^2_{\phi}f\dot{\phi}^2
+3a^3\partial_{\phi}f\ddot{\phi}+9\dot{a}a^2\partial_{\phi}f\dot{\phi}-a^3V\sqrt{1-\dot{\phi}^2}-a^3\rho_s,
\end{eqnarray}
and $\rho_s$ is point-like lagrangian for a perfect fluid (\cite{hawking}).

In this system, besides dark energy, the Universe is composed by matter 
(ordinary and dark) and radiation. Both dark matter and ordinary matter are 
treated as dust, hence represented by the same entity here. As the Universe 
expands, matter's density decrease with $a^{-3}$ while radiation's with 
$a^{-4}$. The lagrangian above contains second-order terms which are more 
tedious to deal with. Since the action limits are fixed, we can integrate these 
terms by parts, without loss of generality, so we can work with a first-order 
lagrangian, which reads
\begin{eqnarray}
\label{pll2}
L&=&-6f\dot{a}^2a-6a^2\dot{a}\partial_{\phi}f\dot{\phi}-a^3V\sqrt{1-\dot{\phi}^2}-\frac{3a^3}{2f}(\partial_{\phi}F\dot{\phi})^2-\rho^0_m-\frac{\rho^0_r}{a},
\end{eqnarray}
where $\rho^0_m$ and $\rho^0_r$ are the recent values of the total density of 
matter and radiation, respectively, in the Universe.

\section{Noether Symmetries}

Our system may now be constrained to that which is endowed with a Noether 
symmetry by finding the forms of $f(\phi)$ and $V(\phi)$ that allow symmetrical 
point transformation. This means that our lagrangian shall have such a form that 
a specific continuous transformation of the generalized coordinates 
$a\rightarrow\bar{a}$ and $\phi\rightarrow\bar{\phi}$ preserves the general form 
of the functional,
\begin{equation}
\label{lpreserve}
L(\bar{a},\bar{\phi})=L(a,\phi).
\end{equation}

In order to find the function forms of $V(\phi)$ and $f(\phi)$ that allow such 
transformation, we need to apply a certain vector field on the lagrangian 
(\ref{pll2}). This vector field, $\mathbf{X}$, is then called a variational 
symmetry, or \emph{complete lift}, and reads
\begin{equation}
\label{vf}
\mathbf{X}\equiv\alpha^i\partial_{q^i}+\dot{\alpha}^i\partial_{\dot{q}^i},
\end{equation}
where the coefficients $\alpha^i$ are functions of the generalized coordinates 
$a,\phi$. The operation of $\mathbf{X}$ on the lagrangian is simply the Lie 
derivative of $L$ along this vector field ($L_{\mathbf{X}}L$). According to the 
Noether theorem, if this derivative vanishes, there will be a conserved quantity 
named \emph{Noether charge}. Hence, this will be a variational symmetry if
\begin{equation}
\label{condition1}
\mathbf{X}L=L_{\mathbf{X}}L=0,
\end{equation}
such that
\begin{equation}
\label{condition2}
L_{\Delta}\left<\theta_L,\textbf{X}\right>=0,
\end{equation}
where $\Delta=d/dt$ is the dynamical vector field and
\begin{equation}
\theta_L=\frac{\partial L}{\partial\dot{q}^i}dq^i,
\end{equation}
is the locally defined Cartan one-form. The brackets represent the scalar 
product between vector field and one-form, in the Dirac notation. Thus, the 
Noether charge reads
\begin{equation}
\label{nt}
\Sigma_0\equiv\left<\theta_L,\textbf{X}\right>=\alpha^i\frac{\partial 
L}{\partial\dot{q}^i}.
\end{equation}

The condition (\ref{condition1}) reads in full form,
\begin{eqnarray}
\label{condition}
0 &=& XL \\ \nonumber
&=& \alpha\partial_aL+\beta\partial_{\phi}L+\left(\dot{a}\partial_a\alpha+\dot{\phi}\partial_{\phi}\alpha\right)\partial_{\dot{a}}L+\left(\dot{a}\partial_a\beta+\dot{\phi}\partial_{\phi}
\beta\right)\partial_{\dot{\phi}}L,
\end{eqnarray}
which for our system becomes

\begin{eqnarray}
\label{nR}
0 &=& 
\alpha\left(-6f\dot{a}^2-12a\dot{a}\partial_{\phi}f\dot{\phi}-3a^2V\sqrt{1-\dot{\phi}^2}-\frac{9a^2(\partial_{\phi}f)^2\dot{\phi}^2}{2f}+\frac{\rho^0_r}{a^2}\right) \nonumber
\\
&+&\beta\left(-6\partial_{\phi}f\dot{a}^2a-6a^2\dot{a}\partial^2_{\phi}f\dot{\phi}-a^3\partial_{\phi}V\sqrt{1-\dot{\phi}^2}+\frac{3a^3(\partial_{\phi}f)^3\dot{\phi}^2}{2f^2}
-\frac{3a^3\partial_{\phi}f\partial^2_{\phi}f\dot{\phi}^2}{f}\right)\nonumber
\\
&+&
\left(\partial_a\alpha\dot{a}+\partial_{\phi}\alpha\dot{\phi}\right)\left(-12f\dot{a}a-6a^2\partial_{\phi}f\dot{\phi}\right)+\left(\partial_a\beta\dot{a}+\partial_{\phi}\beta\dot{\phi}
\right)\nonumber
\\
&\times&\left(-6a^2\dot{a}\partial_{\phi}f+\frac{a^3V\dot{\phi}}{\sqrt{1-\dot{\phi}^2}}-3a^3\frac{(\partial_{\phi}f)^2\dot{\phi}}{f}\right),
\end{eqnarray}

where $\alpha=\alpha^1$ and $\beta=\alpha^2$.

The equation above must hold for any value of $\dot{a}$ and $\dot{\phi}$. If it 
were a polynomial equation for these dynamical variables one could simply make 
all coefficients equal to zero, but the different powers of the square roots 
turn the task a little more complicated. We shall differentiate with respect to 
these quantities and evaluate the resulting equations for different values of 
$\dot{a}$ and $\dot{\phi}$, then we get the solutions for $\alpha(a,\phi)$, 
$\beta(a,\phi)$, $V(\phi)$ and $f(\phi)$. 

Making $\dot{a}=\dot{\phi}=0$ in (\ref{nR}) we get
\begin{equation}
\label{n1}
3\alpha a^2V-\alpha\frac{\rho^0_r}{a^2}+\beta a^3\partial_{\phi}V=0.
\end{equation}

Differentiating equation (\ref{nR}) three times with respect to $\dot{\phi}$ and 
evaluating at $\dot{\phi}=0$ and $\dot{a}=1$ gives
\begin{equation}
3a^3V\partial_a\beta=0,
\end{equation}
hence $\beta\neq\beta(a)$. Similarly, differantitating (\ref{nR}) once with 
respect to $\dot{\phi}$, multiplying it by $(1-\dot{\phi}^2)^{3/2}$ and 
evaluating at $\dot{\phi}=1$ and $\dot{a}=0$ we get
\begin{equation}
V\partial_{\phi}\beta=0,
\end{equation}
and we conclude that $\beta=\beta_0$ is a constant, since the potential must be 
non-zero. The fourth derivative of (\ref{nR}) with respect to $\dot{\phi}$, 
evaluated at $\dot{\phi}=0$ and taking into account that 
$\partial_{\phi}\beta=0$ leads to
\begin{equation}
\label{n2}
9\alpha a^2V+3\beta a^3\partial_{\phi}V=0.
\end{equation}

Since $\rho^0_r\neq0$, dividing (\ref{n2}) by $3$ and equating  with (\ref{n1}) 
results in $\alpha=0$ and $V=V_0$, constant potential. Thus, equation (\ref{nR}) 
reduces to
\begin{equation}
-6\partial_{\phi}f\dot{a}^2a-6a^2\dot{a}\partial^2_{\phi}f\dot{\phi}+\frac{
3a^3(\partial_{\phi}f)^3\dot{\phi}^2}{2f^2}-\frac{3a^3\partial_{\phi}
f\partial^2_{\phi}f\dot{\phi}^2}{f}=0,
\end{equation}
and it is clear that $\partial_{\phi}f=0$. The coupling must then be minimal.

\section{Equations of Motion}

The lagrangian (\ref{pll2}), for constant self-interaction potential and $f=1/2$ 
(to regain Einstein's constant according to the notation adopted), reads
\begin{equation}
\label{lag}
L=-3a\dot{a}^2-V_0a^3\sqrt{1-\dot{\phi}^2}-\rho^0_m-\frac{\rho^0_r}{a}.
\end{equation}

The Friedmann equation is obtained through the energy equation 
$E_L=\dot{a}\frac{\partial L}{\partial\dot{a}}+\dot{\phi}\frac{\partial 
L}{\partial\dot{\phi}}-L$, which gives
\begin{equation}
\label{friedmann}
H^2=\frac{1}{3}\rho,
\end{equation}
where $H=\dot{a}/a$ is the Hubble parameter and $\rho=\rho_m+\rho_r+\rho_{\phi}$ 
is the total energy density of the fields, being 
\begin{equation}
\label{dde}
\rho_{\phi}=\frac{V_0}{\sqrt{1-\dot{\phi}^2}},
\end{equation}
the energy density of the tachyon field, $\rho_m=\rho^0_m/a^3$ and 
$\rho_r=\rho^0_r/a^4$ the matter's and the relativistic material's densities, 
respectively.

The Euler-Lagrange equation for the scalar factor, together with 
(\ref{friedmann}), provides the acceleration equation, being
\begin{equation}
\label{acceleration}
\frac{\ddot{a}}{a}=-\frac{1}{6}\left(\rho+3p\right),
\end{equation}
where $p=p_r+p_{\phi}$ is the pressure of the fields (as usual, matter behaves 
as dust so $p_m=0$), and $p_r=\rho_r/3$. The pressure exerted by the tachyon 
field is
\begin{equation}
\label{pde}
p_{\phi}=-V_0\sqrt{1-\dot{\phi}^2}.
\end{equation}

The Euler-Lagrange equation for the tachyon field gives the generalized 
Klein-Gordon equation for the field, which is the same as the fluid equation for 
dark energy when written in terms of its energy density and pressure
\begin{equation}
\label{kg}
\dot{\rho}_{\phi}+3H\left(\rho_{\phi}+p_{\phi}\right)=0.
\end{equation}

An equation of state in the form $p(\rho)$ can now be written for the tachyon field. From equation (\ref{dde}) we see that $\sqrt{1-\dot{\phi}^2}=V_0/\rho_{\phi}$, which when substituted in (\ref{pde}) yields
\begin{equation}
\label{pdde}
p_{\phi}=-\frac{V_0^2}{\rho_{\phi}}.
\end{equation}

The Chaplygin gas is a fluid described by an equation of state of the kind 
\begin{equation}
\label{chges}
p=-\frac{A}{\rho},
\end{equation}
where $A$ is a positive defined constant, which is precisely the same as (\ref{pdde}) for  $A=V_0^2$. Thus, as widely know from the literature, see e.g. (\cite{frolov,kamenshchik3,chimento,delcampo}), a tachyon field only minimally coupled to the scalar curvature, of constant potential, behaves as a Chaplygin gas.

\section{Noether Constant}

Any lagrangian system endowed with a Noether symmetry will present a constant of 
motion, as stated by Noether's theorem. The Noether charge (\ref{nt}) here 
becomes
\begin{eqnarray}
\label{nt1}
\Sigma_0&=&\alpha\frac{\partial L}{\partial\dot{a}}+\beta\frac{\partial 
L}{\partial\dot{\phi}} \nonumber
\\
&=&\frac{V_0a^3\dot{\phi}}{\sqrt{1-\dot{\phi}^2}},
\end{eqnarray}
which is simply the first integral of (\ref{kg}).

\section{Solutions}

The energy density of the Chaplygin gas, and its pressure, can be rewritten as 
functions of the scale factor, using equation (\ref{nt}). These forms are well 
known from literature and read
\begin{equation}
\label{dpa}
\rho_{\phi}=\sqrt{\frac{\Sigma_0^2}{a^6}+V^2_0};\qquad 
p_{\phi}=-\frac{V_0^2}{\sqrt{\frac{\Sigma_0^2}{a^6}+V^2_0}},
\end{equation}
where $\Sigma_0$ is the Noether constant. From this equation, we see the dual 
nature of the Chaplygin gas, which behaves as dust matter for $a\leq1$
\begin{equation}
\label{chm}
\rho_{\phi}\sim\frac{\Sigma_0}{a^3}; \qquad p_{\phi}\sim0,
\end{equation}
and as a cosmological constant for $a\geq1$
\begin{equation}
\label{chc}
\rho_{\phi}\sim V_0; \qquad p_{\phi}\sim-V_0.
\end{equation}
 
In order to obtain our parameters' curves with respect to the redshift, we use 
the relationship $a=1/(1+z)$. The Friedmann equation (\ref{friedmann}) then 
becomes
\begin{equation}
\label{friedmannz}
H^2=\frac{1}{3}\left(\sqrt{\Sigma_0^2(1+z)^6+V^2_0}
+\rho^0_m(1+z)^3+\rho_r^0(1+z)^4\right).
\end{equation}

The equation above can be written in a dimensionless form by dividing it by the Hubble constant, $H_0^2\equiv H(0)^2=\rho^0/3$, where $\rho^0$ is the current density of all fluids in the Universe, giving
\begin{equation}
\frac{H^2}{H_0^2}=\left(\sqrt{\bar{\Sigma}_0^2(1+z)^6+\bar{V}^2_0}
+\Omega^0_m(1+z)^3+\Omega_r^0(1+z)^4\right),
\end{equation}
where $\Omega_i^0\equiv\rho_i^0/\rho^0$ is the current density parameter of the $i$-th component. The bars indicate that the constants have also been divided by the current density, i.e., $\bar{\Sigma}_0=\Sigma_0/\rho^0$ and $\bar{V}_0=V_0/\rho^0$, then the density parameter for the Chaplygin gas is simply
\begin{equation}
\label{dedp}
\Omega_{\phi}^0=\sqrt{\bar{\Sigma}_0^2+\bar{V}^2_0}.
\end{equation}

This last relationship allows us to investigate the evolution of the Hubble parameter in terms of dark energy's density parameter, instead of the Noether charge. Finally, we write 
\begin{equation}
\label{hf}
H=H_0\left[\sqrt{[(\Omega^0_{\phi})^2-\bar{V}_0^2](1+z)^6+\bar{V}^2_0}
+\Omega^0_m(1+z)^3+\Omega_r^0(1+z)^4\right]^{1/2}.
\end{equation}

Recent observations (\cite{wmap9,planck}) limit the range of values associated to these parameters. In particular, there is great confidence that $\Omega_r^0\sim8,5\times10^{-5}$, so we may adopt this result but will constrain the four remaining parameters (namely $H_0, \Omega_m^0, \Omega_{\phi}^o$, and $\bar{V}_0$) via $H(z)$ data.

\begin{table}[!h]
\centering
\begin{tabular}{|ccc|}
\hline
\hline
z & $H_{obs}$ & $\sigma$  \\ \hline
0.07 & 69 & 19.6 \\ \hline
0.09 & 69 & 12 \\ \hline
0.12 & 68.6 & 26.2 \\ \hline
0.17 & 83 & 8 \\ \hline
0.179 & 75 & 4 \\ \hline
0.199 & 75 & 5 \\ \hline
0.2 & 72.9 & 29.6 \\ \hline
0.24 & 79.69 & 3.32 \\ \hline
0.27 & 77 & 14 \\ \hline
0.28 & 88.8 & 36.6 \\ \hline
0.352 & 83 & 14 \\ \hline
0.4 & 95 & 17 \\ \hline
0.43 & 86.45 & 3.27 \\ \hline
0.48 & 97 & 62 \\ \hline
0.593 & 104 & 13 \\ \hline
0.68 & 92 & 8 \\ \hline
0.781 & 105 & 12 \\ \hline
0.875 & 125 & 17 \\ \hline
0.88 & 90 & 40 \\ \hline
0.9 & 117 & 23 \\ \hline
1.037 & 154 & 20 \\ \hline
1.3 & 168 & 17 \\ \hline
1.43 & 177 & 18 \\ \hline
1.53 & 140 & 14 \\ \hline
1.75 & 202 & 40 \\
\hline
\hline
\end{tabular}
\caption{Observational values for $H(z)$ and their respective errors (\cite{hz,hz1,hz2,hz3,hzc}).}
\label{hzdata} 
\end{table}

Table \ref{hzdata} presents 25 measurements of the Hubble parameter from SNe and gamma-ray burst (\cite{hz,hz1,hz2,hz3,hzc}). The function built for the Hubble parameter (\ref{hf}) depends on the redshift, plus four different parameters. Hence, $H=H(z, H_0, \Omega_m^0, \Omega_{\phi}^0, \bar{V}_0)$. The values assumed by these parameters that best fit the observational data are the ones the minimize the function 
\begin{equation}
\chi^2=\sum\limits_{i=1}^{25}  \left[\frac{H_{obs}(z_i)-H(z_i, H_0,\Omega_m^0, \Omega_{\phi}^0, \bar{V}_0)}{\sigma_i}\right]^2.
\end{equation}

A primary condition for a good fit is that $\chi^2/dof\leq 1$, where $dof$ stands for \emph{degrees of freedom} and in this case is given by the number of data points, $dof=25$. Our minimized $\chi^2$ is given by $H_0=69.6524, \Omega^0_m=0.288261, \Omega_{\phi}^0=0.711654$ and $\bar{V}_0=0.709957$ resulting in $\chi^2=12.8676$ and $\chi^2/dof=0.507504$. Marginalizing over two parameters allows us to analyze the correlation of the remaining by plotting the contours of their distributions within some confidence interval.

\begin{figure}[!h]
\centering
\includegraphics[scale=0.5]{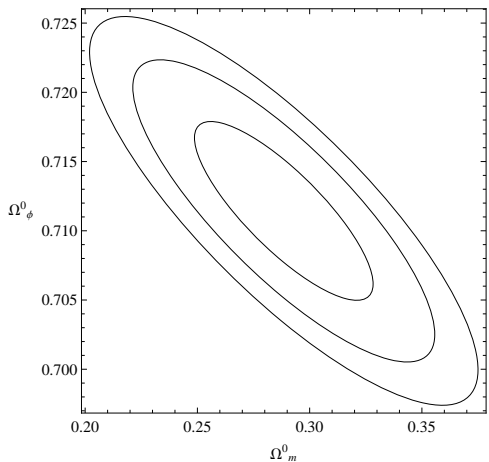}
\caption{Confidence intervals for 1-,2- and 3-$\sigma$ for the density parameters $\Omega_m^0$ and $\Omega^0_{\phi}$.}
\label{figfm}
\end{figure}

\begin{figure}[!h]
\centering
\includegraphics[scale=0.5]{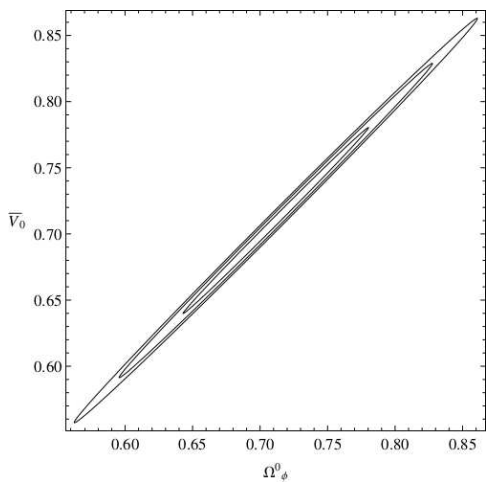}
\caption{Confidence intervals for 1-,2- and 3-$\sigma$ for the density parameter $\Omega^0_{\phi}$ and the constant potential $\bar{V}_0$.}
\label{figfv} 
\end{figure}

The correlation between dark energy and matter density parameter is strong. The $H(z)$ does not seem to impose very strict constraints for our current matter density, but for dark energy we see that within 3-$\sigma$ all points lie in the range $0.697\leq\Omega_{\phi}^0\leq0.726$, Fig.\ref{figfm}. The correlation between $\Omega_{\phi}^0$ and $\bar{V}_0$ is much stronger, as one would expect since the potential defines the energy density. Nevertheless, it is interesting enough to see the form these ellipses take in Fig. \ref{figfv} whilst the current density parameter for dark energy is given by (\ref{dedp}). The case $\Omega_{\phi}^0=\bar{V}_0$ is just the cosmological constant scenario. From the figure, we see very thin ellipses with a slope close to the unity. The best fit parameters listed above show a very small difference between the two of them, and as we will see this difference grows bigger in the past, but there is a high tendency for the cosmological constant.

\begin{figure}[!h]
\centering
\includegraphics[scale=0.5]{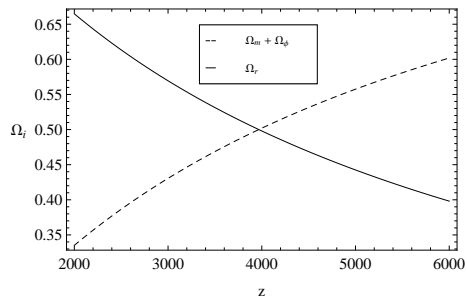}
\caption{Density parameters plotted for high redshift values. The dashed line 
represent the matter fields, where the Chaplygin gas is included once it behaves 
as dust at this point. Solid line stand for radiation's energy density parameter. Equality in densities happen at $z=3968.15$.}
\label{figoh}
\end{figure}

The evolution of the density parameters for different redshift scales are shown below. In Fig. \ref{figos} radiation is neglected for its energy density is too small to be observed. As the redshift increases dark energy falls but ever more slowly, and for values $z\geq2$ its density decreases so smoothly that it almost appears to be constant. This is due to the small difference between $\Omega^0_{\phi}$ and $\bar{V}_0$ that makes the tachyon field dynamical and the Chaplygin gas property thrive, from equations (\ref{dpa}) and (\ref{chm}) it becomes clear that dark energy decays into matter fields as the redshift grows and the term $\bar{\Sigma}_0$ outpaces $\bar{V}_0$. Furthermore, we are now looking at the matter era, hence the almost constat behavior.  In Fig. \ref{figoh} we see the evolution of the density parameters for radiation and the combination of the Chaplygin gas and matter, since the first behaves as the latter. In this scenario, the densities equality happen a bit earlier in out history than expected from the cosmological constant case. For instance, we have $z_{eq}=3968.15$ (approximately 37.5 thousands years since the beginning of the Universe) whereas for a non-dynamical field describing we would expect $z_{eq}\sim3600$ (about 47 thousand years old).
Although a transition happening at higher redshifts does not influence 
the time at which recombination occurs (once it depends on the temperature), the 
sooner increase in dark matter's energy density allows it to combine and form 
potential wells earlier, giving room for structure formation, some of which we have recently discovered and turned out to be quite old (\cite{jkcs1,jkcs2}).

\begin{figure}[!h]
\centering
\includegraphics[scale=0.5]{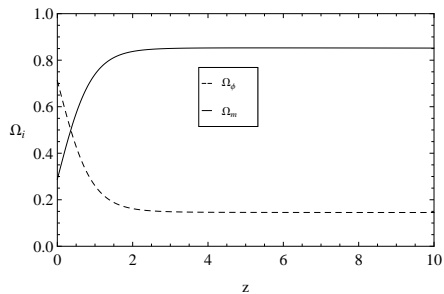}
\caption{The dotted lines stand for dark energy while the solid lines represent 
dark matter. As we enter the matter dominated era, dark energy decays into matter fields contributing even more for its dominance, with its energy density falling ever more slowly, assuming an almost constant behavior.}
\label{figos}
\end{figure}

\begin{figure}[!h]
\centering
\includegraphics[scale=0.5]{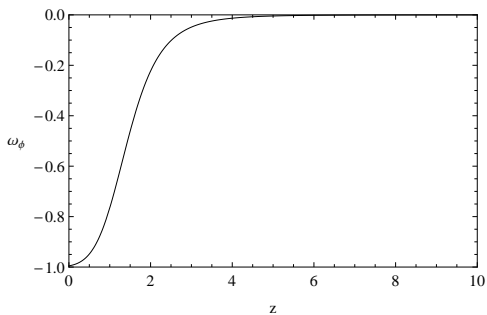}
\caption{Ratio between pressure and energy density for the Chaplygin gas. Any small difference between $\Omega_{\phi}^0$ and $\bar{V}_0$ grows considerably with the redshift and dark energy eventually becomes a pressureless field, hence matter.}
\label{figw}
\end{figure}

The ratio $\omega_{\phi}=p_{\phi}/\rho_{\phi}$, between dark energy's pressure 
and energy density is show in Fig.~\ref{figw}. As expected from equation 
(\ref{dpa}), the ratio tends asymptotically to $-1$ as the Universe expands but approaches zero quickly as the redshift increases, when dark energy finally becomes a pressureless field. It becomes clearer the role of the Chaplygin gas even though we are approaching the cosmological constant in recent times.

\begin{figure}
\centering
\includegraphics[scale=0.5]{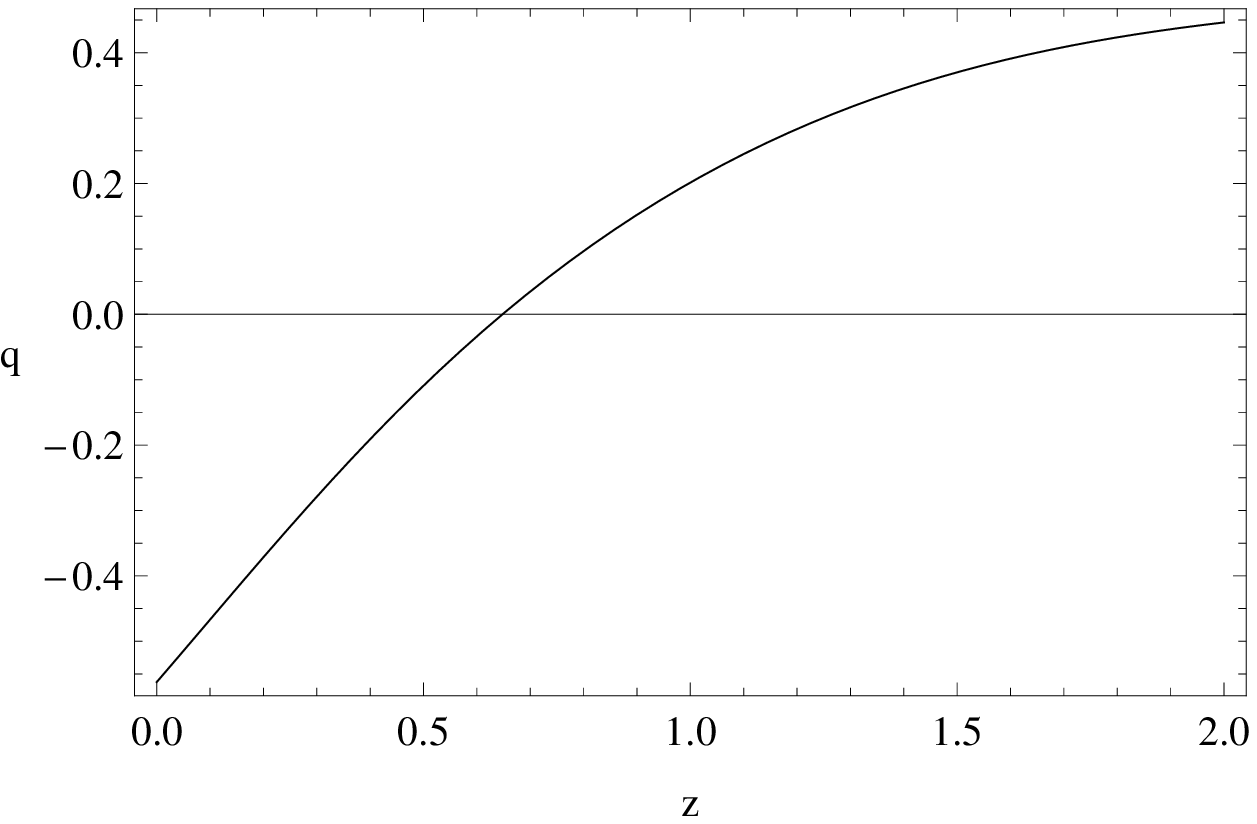}
\caption{Deceleration parameter. Expansion becomes accelerated at $z_t=0.65$ and the current value of this parameters stands at $q_0=-0.56$.}
\label{q}
\end{figure}

The deceleration parameter $q$ is plotted in Fig.~\ref{q}. The transition from a decelerated to an accelerated expansion happens at $z_t=0.65$, while for our current time $q_0=-0.56$, both results in agreement with observations (\cite{qdata}).   

\begin{figure}
\centering
\includegraphics[scale=0.5]{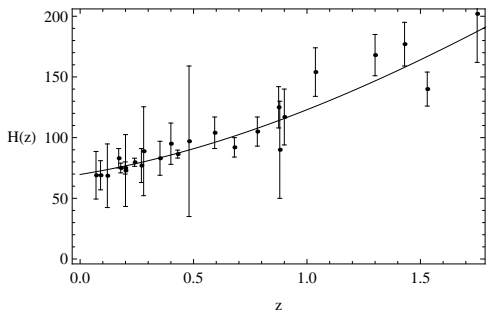}
\caption{The observational data for the Hubble 
parameter (\cite{hz,hz1,hz2,hz3,hzc}) favor the dynamical field over the 
cosmological constant. The more dynamical the tachyon field is, greater is the 
inclination of the $H(z)$ curve. Even with such big error bars, we can identify 
the case $\bar{V}_0=0.69$ as being the one that accommodates the more data.}
\label{fhz}
\end{figure}

Although the Chaplygin gas has been extensively studied before, new observational data provides great motivation to revisit the model and set new constraints. 
The evolution of the Hubble parameter described by this model, together with the data we used to define our parameters is shown in Fig. \ref{fhz}. Unfortunately, there are not satisfactorily many measurements to make solid statistics for this parameter as there are for the distance modulus, for instance. Also, the errors associated with the data from gamma-ray bursts are much bigger than one would desire them to be. Nevertheless, these sources provide information from a much younger Universe compared to SNe, making it worthwhile for testing models and constraining parameters.

\section{Conclusions}

In this work, we started from a general action where a tachyon field represents 
the nature of dark energy. We allowed it to be non-minimally coupled to the 
scalar curvature and we considered the connection to be independent through the 
Palatini approach. Dark matter, baryonic matter and relativistic material were 
included in source fields, as our intention was to build a more complete model. 
Instead of establishing the self-interaction potential and the coupling function 
in a \emph{ad-hoc} manner, we stated that symmetry should play a more primary 
role and only functions capable of composing a continuous and symmetric point 
transformation on the generalized coordinates would be considered. This lead to  
the simpler case where the tachyon field is only minimally coupled and its 
potential is constant, behaving as a Chaplygin gas. 

The theoretical framework of the Chaplygin gas has been deeply investigated and 
is widely found in the literature. For this reason, we focused on more recent 
observational data to 
constrain the dynamics of the system. The Hubble parameter suggests that, be the 
Chaplygin gas the underlying nature of dark energy, it shall be slightly dynamical, 
 as opposed to its cosmological constant particular case since, as we see, any small difference between its constant potential and current density parameter grows considerably with the redshift. As this component presents a dual behavior, acting as 
dark energy for small redshifts and decaying into matter fields later on, the 
matter era begins earlier 
in the history of the Universe, what could help explain older structures.

\section{Acknowledgments}

The authors would like to thank CNPq for the financial support that allowed this work to be done. Also, we are grateful for the pertinent suggestions made by the Referee.

\bibliography{biblio}
\end{document}